\documentclass[useAMS,usenatbib]{mn2e}
\usepackage{graphicx}
\usepackage{color}
\usepackage{url}
\usepackage{longtable}

\definecolor{myred}{rgb}{0.7,0.0,0.2}

\definecolor{myblue}{rgb}{0.0,0.2,0.7}

\definecolor{mygreen}{rgb}{0.2,0.7,0.0}

\title[O star spin rates]{The spin rates of O stars in WR + O Magellanic Cloud binaries}
\author[M. Shara et al.]{Michael M. Shara$^{1}$\thanks{E-mail: mshara@amnh.org}, Steven M. Crawford$^{2,3}$, Dany Vanbeveren$^{4}$, 
\newauthor {Anthony F. J. Moffat$^{5}$, David Zurek$^{1}$ and Lisa Crause$^{2}$}
\\
$^{1}$Department of Astrophysics, American Museum of Natural History, Central Park West at 79th Street, New York, NY 10024, USA\\
$^{2}$South African Astronomical Observatory, P.O. Box 9, Observatory 7935, Cape Town, South Africa\\
$^{3}$Space Telescope Science Institute, 3700 San Martin Drive, Baltimore, MD 21218, USA\\
$^{4}$Astrophysical Institute, Vrije Universiteit Brussel, Pleinlaan 2, 1050, Brussels, Belgium\\
$^{5}$D\'epartement de Physique, Universit\'e de Montr\'eal, CP 6128 Succ. C-V, Montr\'eal, QC H3C 3J7, Canada\\
}

\begin{document}

\date{Accepted  Received }


\maketitle


\begin{abstract}
Some massive, merging  black holes (BH) may be descendants of binary O stars. The evolution and mass transfer between these O stars determines the spins of their progeny BH. These will be measurable with future gravitational wave detectors, incentivizing the measurement of the spins of O stars in binaries. We previously measured the spins of O stars in Galactic Wolf-Rayet (WR) + O binaries. Here we measure the $v_e sini$ of four LMC and two SMC O stars in WR + O binaries to determine whether lower metallicity might affect the spin rates. We find that the O stars in Galactic and Magellanic WR + O binaries display average $v_e sini$ = 258 $\pm$ 18 km/s and 270 $\pm$ 15 km/s, respectively. Two LMC O stars measured on successive nights show significant line width variability, possibly due to differing orbital phases exhibiting different parts of the O stars illuminated differently by their WR companions. Despite this variability, the $v_e sini$ are highly super-synchronous but distinctly subcritical for the O stars in all these binaries; thus we conclude that an efficient mechanism for shedding angular momentum from O stars in WR + O binaries must exist.  This mechanism, probably related to Roche lobe overflow-created dynamo magnetic fields, prevents nearly 100\% breakup spin rates, as expected when RLOF operates, as it must, in these stars. A Spruit-Tayler dynamo and O star wind might be that mechanism.

\end{abstract}



\section{Introduction}

A detailed motivation for determining the spin speeds of O stars in Wolf-Rayet (WR) + O binaries has been been given in \citet{sha17}, hereafter Paper I. Here we present a concise overview of that motivation, an important limitation to Paper I, and an outline for this paper which seeks to remove that limitation.

\citet{mae00, heg00, hir04, yoo05, bro11} and \citet{eks12} have shown that rapid initial rotation on the main sequence (i.e. equatorial velocities $>$ 200-300 km/s) dramatically changes the evolution and deaths of massive stars. Ultra-luminous supernovae and long-duration gamma-ray bursts \citep{woo06,geo09} may result from the final collapses of massive, rapidly spinning stars. Merging binary black holes (BH) \citep{abb16} originating in WR + O binaries may lead to BH spins that will be measurable with Advanced LIGO and Virgo \citep{pur16}. Those spins are determined by the evolution of the binary components, driven by mass and angular momentum transport between the components. It is clearly desirable to measure both massive stars' spin rates in binaries, to test theoretical binary evolution calculations and to constrain their predictions of BH spin rates during mergers.
  
\subsection{Single O stars, Binaries and RLOF} 

\citet{pen96} and \citet{how97} measured {\it single} Galactic massive stars' rotation rates, while the VLT Tarantula survey examined similar LMC stars \citep{duf13, ram13}. The majority of these stars in both galaxies display a modest average equatorial velocity $v_e \sim$100 km/s, but these are all single O-stars on or near the Main Sequence. A small but significant fraction of them, however, display $v_e > $ 200 km/s, with a few LMC stars achieving 500-600 km/s.  Some or all of the rapid rotators which appear single may be merged binaries with rapid spins. \citet{vdh93, van98}, and \citet{vaa98} predicted that many or most massive stars must be close binaries. This prediction has now been observationally confirmed \citep{mas98, mas09, san11, san13}. Population synthesis modeling which includes binaries has reproduced the observed distribution of rotation rates of massive stars \citep{dmk13}.

The most massive primaries (up to $\sim$ 100 $M_{\odot}$) in binaries may lose much of their initial masses via stellar winds, greatly increasing the binary periods. WR + O binaries with periods of days are quite common, and they demonstrate that Roche lobe overflow (RLOF) and/or common envelope (CE) phases must have occurred earlier in these stars' lives \citep{vnn98}. RLOF can lead to rapid rotation of the mass gainer as it accretes mass and angular momentum \citep{pac81,van98}. If this phase is followed by a supernova explosion of the initially more massive primary (which will explode first, even if it is the less massive star at the time) which disrupts the binary, a runaway star \citep{bla61} which is rapidly rotating may result. An example of a runaway star is the early-type supergiant $\zeta$ Pup \citep{van12}, likely ejected from a binary after the RLOF spin-up AND the supernova explosion of the original primary to a runaway BH in the opposite direction, which displays $v_e$ sini of 220 km/s. The inclination i of $\zeta$ Pup has recently been claimed to be $\sim$ 23 deg \citep{ram18} with the star's equatorial velocity being 550 km/s, in excess of 80\% of its critical rotation speed. Less extreme equatorial velocity values \citep{how19} corresponding to a differing interpretation of periodicity derived from \citet{ram18}'s satellite photometry still imply such high rotation that the star must originally have been part of a binary.  

In Paper I we measured the observational line width parameter $v_e sin i$ for the O star in 8 WR + O binaries from line broadening of helium lines to determine whether RLOF had been operative in spinning them up in the past \citep{van98, pet05}. Remarkably, super-synchronous spins ($v_e sin i > $100 km/s in binaries with periods shorter than about 10 days) were found in every O star in the sample, and two more from the literature, strongly indicative of RLOF.

All of the binaries investigated in Paper I are Galactic with approximately solar metalicity. O Stars with lower metalicity and consequent lower atmospheric opacity might shed mass and angular momentum at lower rates than their Galactic counterparts, leading to larger  $v_e$ sini. To check if super-synchronous spins are also present in Magellanic O stars in WR + O binaries, we have therefore now expanded our survey to include four WR + O binaries in the Large Magellanic Cloud, and two in the Small Magellanic Cloud. We note that independent measurements of $v_e sini$ of 5 binary WR + O stars in the SMC have recently been presented by \citet{she16}, which complement and act as independent checks of our results; we discuss them below after presenting our own results.
 
In Section 2 we describe the data and their reductions, and present the high resolution spectra of the helium line of the O stars we study. These stars' derived projected rotation rates are given in section 3. In section 4 we discuss the implications of our results for the overall evolution of rotational velocities in massive binaries, and we briefly summarize our results in Section 5.

\section{Observations and Data Reductions}\label{obs}
\subsection{Observations}

Observations\footnote{Observations were taken under SALT Proposal Code:
2016-2-SCI-055}  of the six target Magellanic stars were obtained with the High Resolution
Spectrograph \citep{cra14} (HRS) of the Southern African Large
Telescope (SALT). A full description of the SALT HRS data reductions is given in Paper I.
In summary, we note that all Magellanic WR+O star spectra were obtained during November 2016 in the low resolution mode of HRS with
a 2.23" arcsec diameter fiber to provide a spectrum over the spectral range of 3700-5500 \AA. 
A single ThAr arc and spectral flats were also obtained in this mode for the purposes of
calibration. All spectra had R $\sim$12,700 and signal-to-noise ratio S/N per pixel $>$ 150,
yielding an instrumental velocity resolution (from measurements of the He I $\lambda$4922 line 
of the slowly rotating B0.2 V star $\tau$ Sco, shown in Figure 1, and as reported in Paper I) 
of  $24\pm 3 $ km/s. A spectrum was also taken, with identical setup, of the well-studied, rapidly rotating O star
 $\zeta$ Oph on 1 June 2016. A list of targets and log of observations is provided in Table 1.

\subsection{HeI and HeII lines}

In Paper I we followed the procedure outlined in detail by \citet{ram13} and \citet{ram15} 
to determine the value for $v_e sin i$ for the stars in our sample. This involved measuring  
the FWHM of the lines in our stars via Voigt profile fitting and continuum subtraction. From the fits,
we determined the FWHM for each of the lines and then converted these to velocities
based on the relationships in Table 1 of \cite{ram15}. This process yielded He II $v_e sin i$ 
values considerably and consistently smaller than those determined for the He I line. We suggested in Paper I that a 
possible explanation for this discrepancy might be due to oblateness and gravity darkening in the O stars. \citet{ree18} 
have criticized this result, showing that careful choice of pseudo-continuum normalization removes the apparent discrepancy. In particular,
instead of approximating the pseudo-continua by low-order polynomials \citep{sha17}, \citet{ree18} fitted Hermite splines to continuum points 
selected by eye, better fitting the WR emission-line structure. Their measurements of HeI $\lambda$4922 are in good agreement with those of \citet{sha17}, but their HeII $\lambda$4541 FWHM values are systematically smaller, by up to almost a factor of two.

\subsection{$\zeta$ Oph's HeI and HeII lines}

The ubiquitous and large differences between the He I and He II velocities of Paper I, and especially the critique of \citet{ree18} have prompted us to check the 
\citet{ram13} and \citet{ram15} methodology via the extensively observed, rapidly rotating O star $\zeta$ Oph. A dozen measurements of $v_e sini$ for $\zeta$ Oph have been published in the past four decades \citep{con77,vog83,sto87,pen96,pul96,how97,bal99,jan00,how01,fre05,sim14,caz17}, ranging from 337 to 400 km/s, and averaging 366 km/s.

The HeI line of $\zeta Oph$ that we observed with SALT/HRS is shown in Figure 2. The FWHM measurement and analysis methodology of \citet{ram15} yield 
$v_e sini$ = $397 \pm 28$ km/s from the He I $\lambda$4922 line and $252 \pm 34$ km/s from the He II $\lambda$4541 line. The agreement of our He I $v_e sin i$ with the dozen authors noted above is encouraging, but it is clear that the He II $v_e sin i$ value determined with the \citet{ram15} formalism is incorrect. In light of the discrepant HeII velocities we withdraw our suggestion from paper I that these stars' HeI and HeII velocities are very different, possibly because of gravity darkening. We will use only the relatively strong He I line in our spectra - that of $\lambda$4922 - in this paper to derive the projected spin velocities of six Magellanic O stars, and re-derive the projected spin velocities of five Galactic O stars.  We use the simple methodology outlined by \citet{zeh18} to derive  $v_e sin i$ = c x (LW/2) / $\lambda_0$, where c is the speed of light and $\lambda_0$ is the rest wavelength of the line being used, followed by a correction for the compound structure (see Figure 1) of the HeI $\lambda$4922 line \citep{mih75}.

\subsection{Results: The O stars' HeI line and the \citet{zeh18} methodology}

Voigt and Gaussian profiles were fit to each HeI line. 
The line profiles and continuum were then fit simultaneously using the
equation:
$$ f(\lambda) = V(\lambda) + C(\lambda)$$
where $V(\lambda)$ is the Voigt Profile which is a second order
polynomial of the form, $C(\lambda) = a + b \lambda + c \lambda^2$.
The data were then normalized by $C(\lambda)$ such that the continuum
would have a value of 1 as displayed in Figures 1 through 13. Next the
noise of the   continuum was estimated by determining the standard
deviation of the observed flux outside the fitted line.
The Full Width at Zero Intensity was then determined by measuring the
high and low wavelengths at which $V(\lambda)$ became
indistinguishable from the noise. 

Adopting the same methodology as \citet{zeh18}, the line widths LW were determined from the wavelengths at which the fitted line profiles 
reached the noise level where the continuum flux had a value of 1. The derived line widths LW with Voigt and Gaussian fits were nearly identical. 
The Voigt profile fits to each of the continuum-divided absorption lines of HeI in $\zeta$ Oph, and in the O stars of each of our five Galactic and six Magellanic WR + O binaries 
are presented in Figures 2 through 13. Our derived $v_e sin i$ of the He I $\lambda$4922 line of $\zeta$ Oph is $397 \pm 8$ km/s, in reasonable agreement with the average of the dozen determinations, 366 km/s, noted above. 

The multi-component nature of the HeI $\lambda$4922 line (see Figure 1) broadens that HeI line by 
$\sim 10 \%$ at 400 km/s and by $\sim 15-20\%$ at $\sim  200$ km/s \citep{sha17}. Microturbulence and macroturbulence of up to 50 km/s \citep{sun13} also broaden the line, but only by $\sim 1-3 \%$ in the 200-400 km/s range, hence we ignore these effects. A reasonable correction is obtained by decreasing the  $v_e sin i$ derived from Voigt profile LW measurements by $10 \%$  at 400 km/s, $20 \%$  at 200 km/s, and extrapolating linearly between those velocity extremes. The key result of this paper - that O stars in WR + O binaries always spin highly super-synchronously but subcritically - is robustly obtained even if we halve or double the above assumption regarding the correction to $v_e sin i$ due to the composite nature of the $\lambda$4922 line.

\newpage

\begin{figure}
\centerline{\includegraphics[width=0.99\columnwidth]{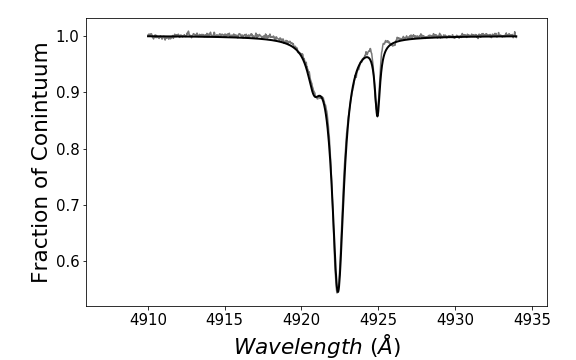}}
\caption{The HeI 4922 absorption line of the O star $\tau\ $Sco on 07 August 2015. In this and all other figures, the solid black line is a three-component model of the HeI 4922 absorption line, while the gray curve
is the HRS data. }\label{spectra}
\end{figure}

\begin{figure}
\centerline{\includegraphics[width=0.99\columnwidth]{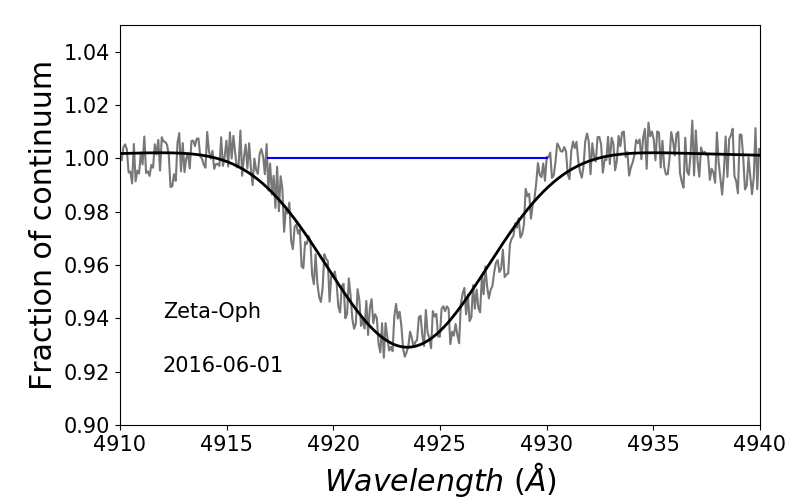}}
\caption{The HeI 4922 absorption line of the O star $\zeta\ $Oph on 01 June 2016. In this and all further figures a horizontal blue line indicates the line width LW as discussed in the text.}\label{spectra}
\end{figure}

\newpage

\begin{figure}
\centerline{\includegraphics[width=0.99\columnwidth]{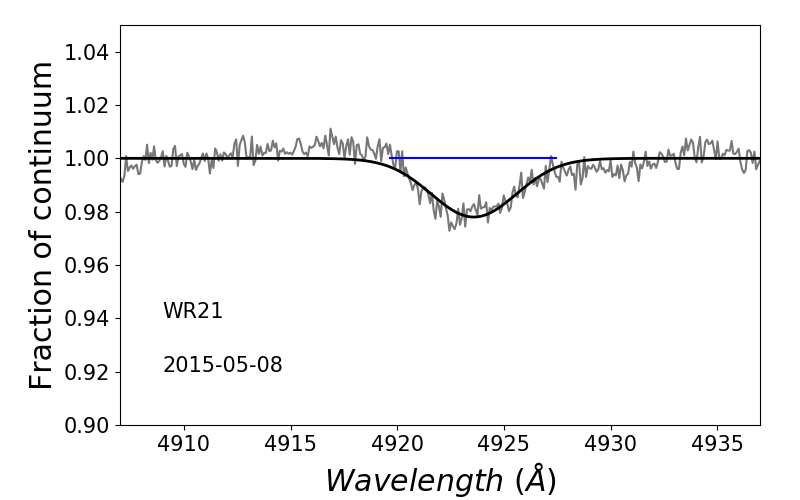}}
\centerline{\includegraphics[width=0.99\columnwidth]{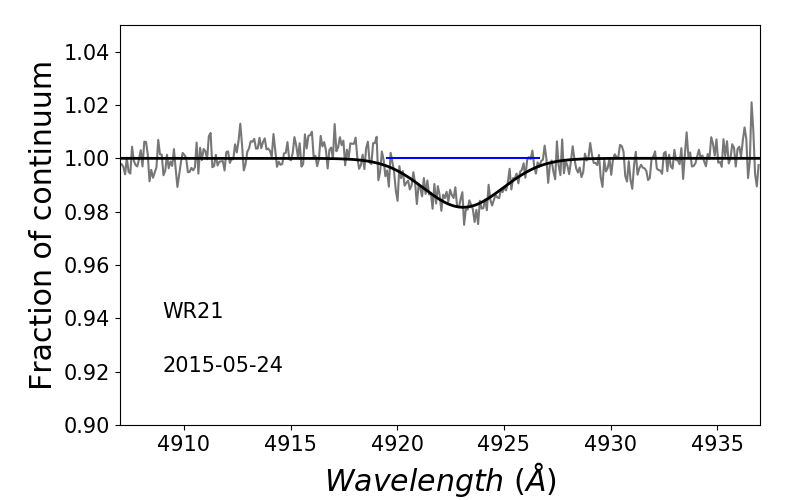}}
\caption{(Top) The HeI 4922 absorption line of the O star in the Galactic WR+O binary WR31 on 08 May 2015.
(Bottom) Same as above but on 24 May 2015.}\label{spectra}
\end{figure}

\begin{figure}
\centerline{\includegraphics[width=0.99\columnwidth]{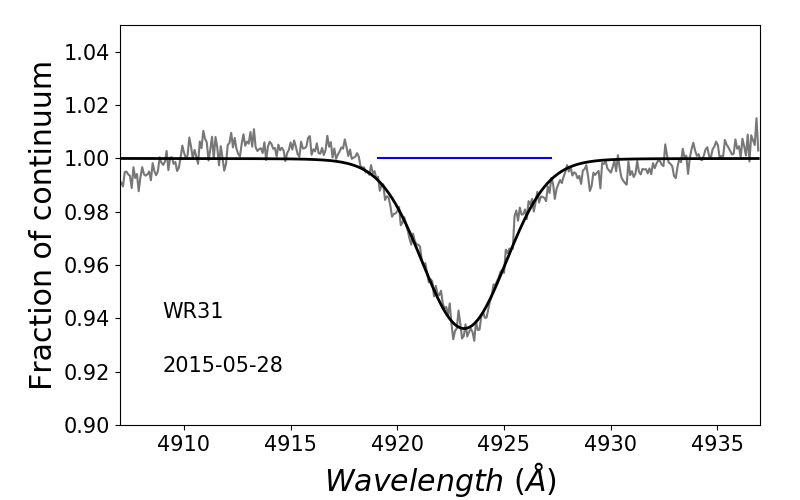}}
\caption{The HeI 4922 absorption line of the O star in the Galactic WR+O binary WR31 on 28 May 2015.}\label{spectra}
\end{figure}

\begin{figure}
\centerline{\includegraphics[width=0.99\columnwidth]{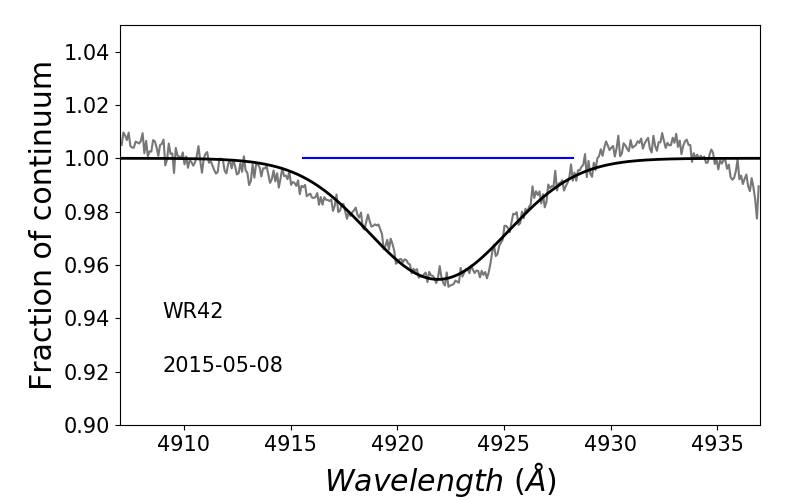}}
\caption{The HeI 4922 absorption line of the O star binary AB7 in the SMC on 03 November 2016.}\label{spectra}
\end{figure}

\begin{figure}
\centerline{\includegraphics[width=0.99\columnwidth]{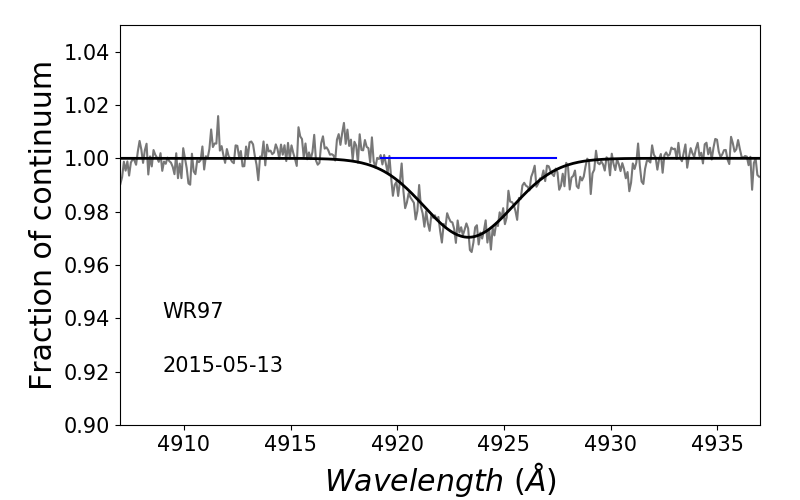}}
\centerline{\includegraphics[width=0.99\columnwidth]{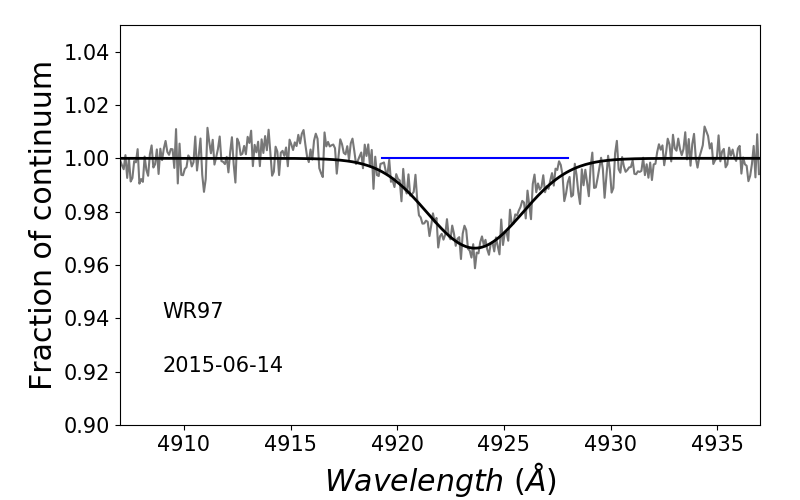}}
\caption{(Top) The HeI 4922 absorption line of the O star in the Galactic WR+O binary WR97 on 13 May 2015.
(Bottom) Same as above but on 14 June 2015.}\label{spectra}
\end{figure}

\begin{figure}
\centerline{\includegraphics[width=0.99\columnwidth]{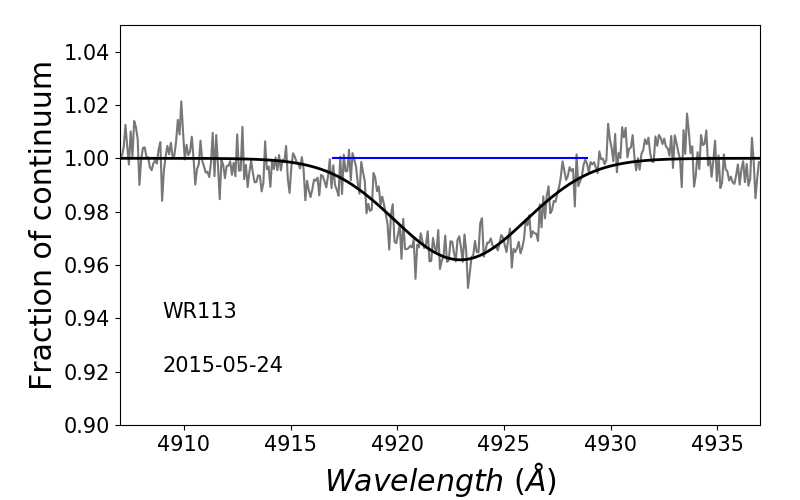}}
\centerline{\includegraphics[width=0.99\columnwidth]{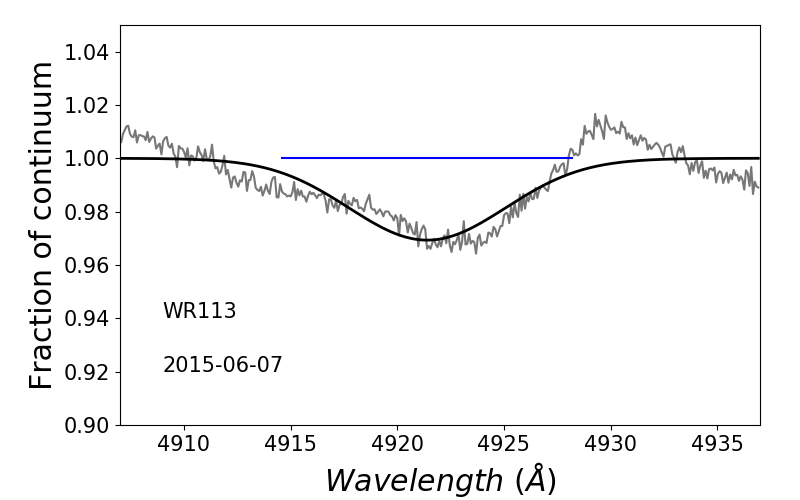}}
\caption{(Top) The HeI 4922 absorption line of the O star in the Galactic WR+O binary WR113 on 24 May 2015.
(Bottom) Same as above but on 07 June 2015.}\label{spectra}
\end{figure}

\begin{figure}
\centerline{\includegraphics[width=0.99\columnwidth]{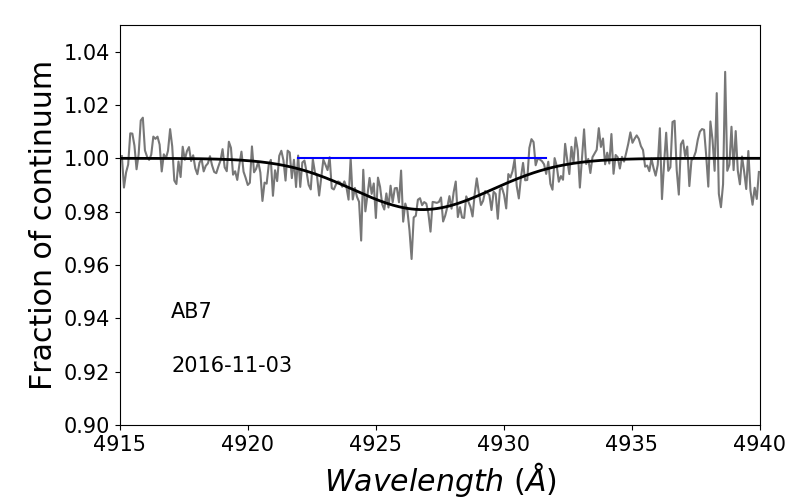}}
\caption{The HeI 4922 absorption line of the O star binary AB7 in the SMC on 03 November 2016.}\label{spectra}
\end{figure}

\begin{figure}
\centerline{\includegraphics[width=0.99\columnwidth]{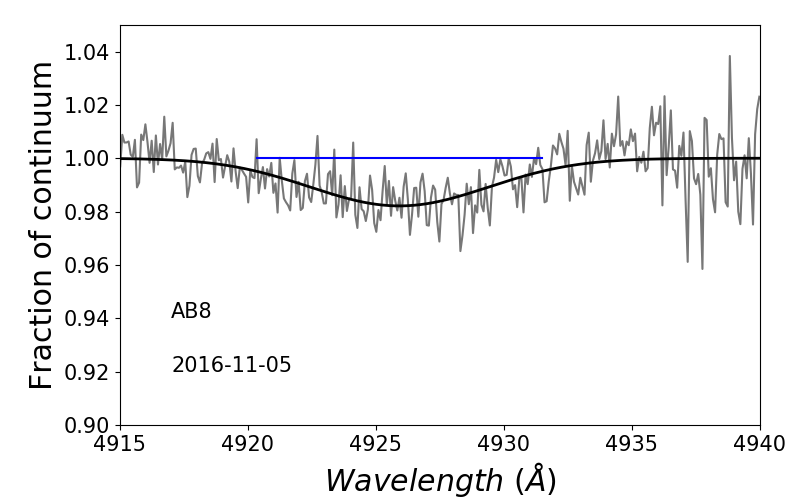}}
\caption{The HeI 4922 absorption line of the O star binary AB8 in the SMC on 05 November 2016.}\label{spectra}
\end{figure}

\begin{figure}
\centerline{\includegraphics[width=0.99\columnwidth]{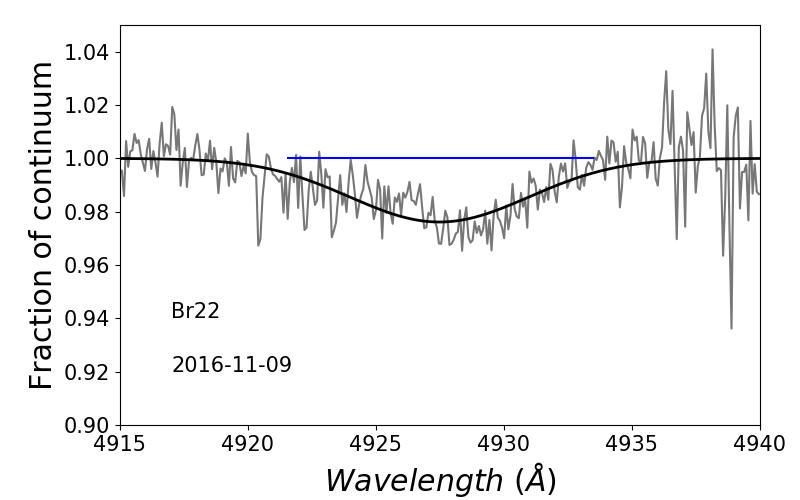}}
\caption{The HeI 4922 absorption line of the O star binary Br22=BAT99-28 in the LMC on 09 November 2016.}\label{spectra}
\end{figure}

\begin{figure}
\centerline{\includegraphics[width=0.99\columnwidth]{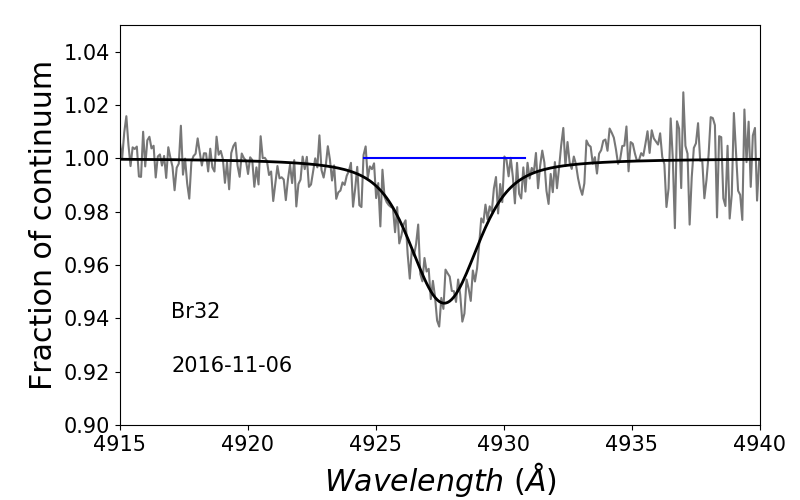}}
\caption{The HeI 4922 absorption line of the O star binary Br32=BAT99-39 in the LMC on 06 November 2016.}\label{spectra}
\end{figure}

\newpage

\begin{figure}
\centerline{\includegraphics[width=0.99\columnwidth]{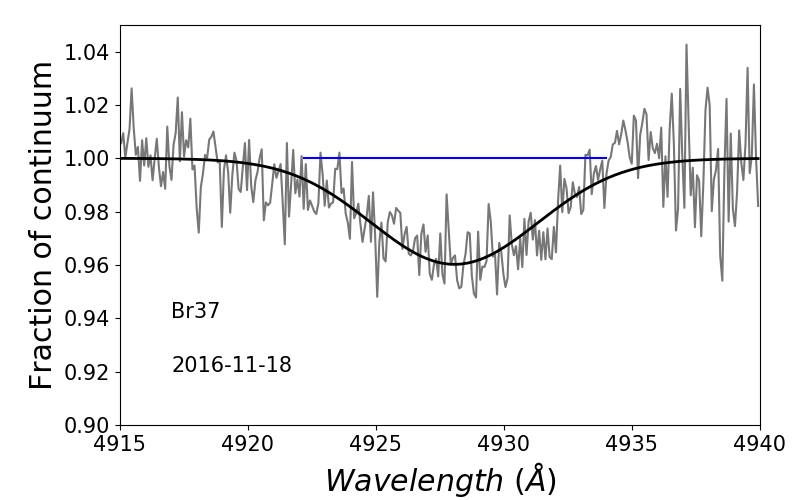}}
\caption{The HeI 4922 absorption line of the O star Br37=BAT99-43 in the LMC binary on 18 November 2016.}\label{spectra}
\end{figure}

\begin{figure}
\centerline{\includegraphics[width=0.99\columnwidth]{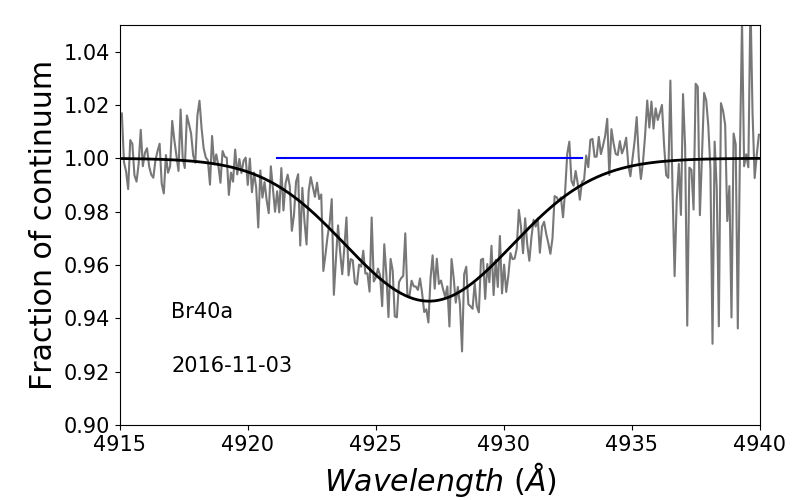}}
\centerline{\includegraphics[width=0.99\columnwidth]{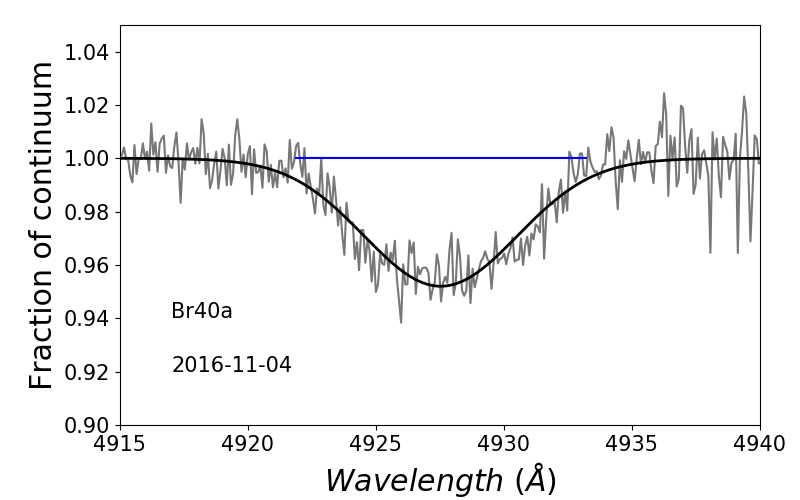}}
\caption{(Top) The HeI 4922 absorption line of the O star binary Br40a=BAT99-49 in the LMC on 03 November 2016. 
(Bottom) Same as above but on 04 November 2016.}\label{spectra}
\end{figure}

\newpage

In Table 1 we report our measured HeI LW and velocities for each of five Galactic and
six Magellanic O stars in WR + O binaries. The smallest LW measured in any of the target stars 
in our sample is $6.32 \pm 0.59 \AA$  which corresponds to a $v_esin i$ of $153 \pm 18 \ $km/s for BAT99-39. 
The largest LW measured is $13.62 \pm1.94 \AA$ which corresponded to a $v_esin i$ of $373 \pm 59 \ $km/s for WR113.

\begin{table*}
 \centering
  \caption{Measured FWHM and $v_e sini$ of Milky Way (MW), LMC and SMC O stars in O+WR star binaries}\label{Tid}
  \begin{tabular}{@{}rllllllll@{}}
  \hline
Star&galaxy&Spectral Types& Period&Observation&He I $\lambda 4922$&HeI $v_e sin i$&$vsin i (Shenar)$\\ &&&(days)&date&LW (\AA)&(km/s)&(km/s)\\
 
\hline
$\zeta$ Oph & MW&O9.5V&--&20160601 &  $13.05 \pm 0.25$ &  $357 \pm 8$&--\\
\\
WR21 &MW&WN5o + O4-6 & 8.3 & 20150508 & $7.80 \pm 0.82$ &  $194 \pm 25$&--\\
WR21 &MW&WN5o + O4-6 & 8.3 & 20150524 & $7.14 \pm 1.24$ &  $175 \pm 38$&--\\ 

WR31 &MW&WN4o + O8V &4.8 & 20150528 & $8.12 \pm 0.71$ &  $203 \pm 22$ &-- \\

WR42 &MW&WC7 +O7V&7.9& 20150508 &  $12.63 \pm 0.68$ &  $344 \pm 21$ & -- \\

WR97 &MW&WN5b + O7&12.6& 20150513 & $8.21 \pm 0.58$ &  $206 \pm 18$  & -- \\
WR97 &MW&WN5b + O7& 12.6&20150614 & $8.70 \pm 0.78$ &  $221 \pm 24$ & -- \\

WR113 &MW&WN5b + O7&12.6& 20160524 & $11.90 \pm 2.56$ &  $319 \pm 78$ & -- \\
WR113 &MW&WN5b + O7&12.6& 20160607 &  $13.62 \pm 1.94$ &  $373 \pm 59$ & -- \\
\\
AB7&SMC&WN4+O6I(f) & 19.2 & 20161103 & $9.69\pm 1.02$ & $250 \pm 31$&$150 \pm 30$ \\
AB8&SMC&WO3+O4V & 16.6 & 20161105 &  $11.17 \pm 1.03$ & $296 \pm 31$& $120 \pm 20$ \\
\\
BAT99-28=Br22&LMC&WC4+O5-6V-III(+O) & 14.9 & 20161109 &  $11.91\pm 1.39$ & $319 \pm 42$&-- \\
BAT99-39=Br32&LMC&WC4+O6V-III(+O) & 1.9 & 20161106 &  $6.32 \pm 0.59$ & $ 153 \pm 18$&-- \\
BAT99-43=Br37&LMC&WN4o+OB  & 2.8 & 20161118 &  $11.83 \pm 1.93 $ & $317 \pm 59 $ &-- \\
BAT99-43=Br37&LMC&WN4o+OB  & 2.8 & 20161124 &  $10.10 \pm 0.66 $ & $262 \pm 20 $&-- \\
BAT99-49=Br40a&LMC&WN4:b+O8V & 31.7 & 20161103 &  $11.91 \pm 1.70$ & $319 \pm 52$&280\\
BAT99-49=Br40a&LMC&WN4:b+O8V & 31.7 &20161104 &  $11.33 \pm 1.71$ & $301 \pm 52$&280\\

 \\

\hline

\end{tabular}\\

Spectral Types and orbital periods from \citet{bre99} = BAT99, \citet{bar01,bas01}, \citet{foe03,fof03}, \citet{stn05}, \citet{sch08}, and \citet{she16}\\
\end{table*}

In Paper I we were able to glean, from the literature, orbital inclinations for all the Galactic WR + O binaries we measured. The Galactic WR stars of Paper I are, of course, more than an order of magnitude closer than the Magellanic binaries we consider here. A strong warning is provided by the SMC star AB6, which is at least quadruple (WR+O and O+O) \citep{she18}. It is likely that at least some of our Magellanic sample stars have unresolved companions, so their masses and orbital inclinations are much less well determined than those of Galactic WR binaries. For this reason we present only $v_e sini$, and do not attempt to deduce equatorial velocities for the Magellanic O stars we consider. 

\section{Rotation speeds}

The average $v_e sin i$ determined from LW for our sample of six Magellanic O stars with measured HeI lines is 270 $\pm$ 15 km/s, which is the projected speed we adopt as representative of O stars in Magellanic WR + O binaries. For 5 Galactic O stars with $v_e sini$ determined from the same HeI line and LW methodology, we find an average projected rotation speed of  258$\pm$ 18 km/s. The mean value of sin i for a random ensemble of stars is $\pi/4$ = 0.7854 (corresponding to i = 51.76 deg), so that the average de-projected rotation speed of our six Magellanic (five Galactic) O stars is 344 (328) km/s. As already noted, we have ignored micro- and macroturbulence, and corrected for the multiple nature of the $\lambda$4922 line.

The synchronous rotation speeds for 11 Galactic WR + O binaries \citep{sha17} with well-determined masses and orbits average to 60 km/s. All but one are observed to be rotating at highly supersynchronous speed. Two of the LMC binaries of our sample, BAT99-39 and BAT 99-43 display such short orbital periods (1.9 and 2.8 days, respectively) that we expect them to be tidally locked. The orbital periods of the four other Magellanic binaries of this paper average 20.6 days, similar to those of the 11 Galactic binaries which average 17.2 days. The 270 km/s average  of the Magellanic O stars is very supersynchronous. The average critical rotation speed for the 11 Galactic O stars noted above is 555 km/sec, and we expect a similar result for the Magellanic O stars. In that case, we find that the Magellanic (Galactic) O stars are spinning with, on average, 270/555 (258/555) = 49 (46) \%  of those stars' critical rotation speeds. This is the key finding of our study. This key result is robust even if micro- and/or macroturbulence, and/or the presence of unresolved companions changes the spin speed by 50-100 km/s. That these supersynchronous speeds are a consequence of binary interaction follows from a comparison of the average $v_e sin i$ of O stars in binaries noted in the previous paragraph with the observed rotational velocities of single LMC and SMC O stars. The distribution of $v_e sin i$ shows a peak at $\sim$ 80 km/s for single O stars in the 30 Doradus region of the LMC \citep{ram13}, and a similar behavior for single O stars in the SMC \citep{mok06}.

Our average measured value $v_esin i$ = 310 km/s for BAT99-49 is in excellent agreement with the value of 280 km/s measured by \citet{she18}. However our measured values for AB7 and AB8 are 250 and 296 km/s, respectively, while those of \citet{she18} are 150 and 120 km/s. It is possible that our estimated spin speed errors are significantly larger than our simple analysis indicates, as Shenar's model atmosphere models use multiple lines to yield spin speeds.  In addition, the presence of more than one unresolved O star on our HRS slit could easily increase the LW of the HeI line we measure, and erroneously increase the spin speed we deduce. It is also clear, as seen in the case of BAT99-43 that significant night-to-night variations in measured LW are observed, possibly due to orbital phase variable illumination of the O star. Ideally, future attempts to measure $v_e sin i$ of O stars in WR + O binaries should include measurement with good orbital phase coverage. {\it Despite these caveats, it seems inescapable that the large $v_e sin i$ of Galactic and Magellanic O stars in WR + O binaries are strongly supersynchronous but also clearly subcritical}.

\section{Tides and RLOF}

Angular momentum transfer accompanies mass transfer during RLOF, wherein some of the mass lost by a donor star is accreted by its companion. This forces the mass gainer to spin-up. When the RLOF-process in a case B binary (i.e. RLOF starts while the mass-loser is hydrogen shell burning) is quasi-conservative, the mass gainer is quickly spun up to its critical Keplerian speed \citep{pac81}. The observed rotational speeds of the O-type companions in the Galactic and Magellanic WR binaries in which it has now been measured are highly super-synchronous, suggesting that mass transfer and spin-up have played important roles during the progenitor evolution. RLOF transfers angular momentum quickly enough to spin up the O-type mass gainers to critical rotation in much less than the evolutionary timescale of a Roch-lobe filling O star, which is of the order of a few hundred thousand years. That the observed spin speeds of the O stars are $\sim 50\%$ of their breakup speeds suggests that an angular momentum-shedding mechanism exists to limit the stars' spinup, or to rapidly decrease it as critical rotation is approached.
One suggestion \citep{van18} is that if, during the mass transfer phase of a massive binary, a Spruit-Taylor dynamo can generate a magnetic field of the
order of a few kG, then rapid spin-up of the mass gainer can be compensated at the expense of a moderate mass loss from the binary. The enhanced dynamo-created magnetic field would be short-lived and difficult to catch during the brief RLOF phase itself.

\section{Conclusions}\label{conclusions}
Theory predicts that the O stars in WR + O binaries must have accreted significant amounts of angular momentum during RLOF from their companions.  We report $v_e sin i$ measurements for six Magellanic Cloud O stars and for five Galactic O stars in WR+O binaries. The average projected equatorial rotational value is 270 km/s for six Magellanic O stars, ranging from 153 to 319 km/s, from the line width of the HeI 4922 line. The average projected equatorial rotational value is 258 km/s for five Galactic O stars, ranging from 185 to 373 km/s. These values are strongly super-synchronous, agreeing with the predictions of short period, massive binary evolution models which include angular momentum transfer during RLOF. Since the rotation speeds are only $\sim$ half the critical speed, a mechanism must also exist to shed angular momentum from the O stars so that they either do not reach critical rotation speeds, or rapidly reduce those speeds by quickly shedding angular momentum. This might be a Spruit-Taylor dynamo whose associated kG magnetic field carries away angular momentum in a wind. 
\section*{Acknowledgments}

The High Resolution Spectrograph (HRS) of the Southern African Large Telescope (SALT) produced the data reported in this paper.
We gratefully acknowledge the fine support of the astronomers and operators at the SALT Observatory. MMS thanks Tomer Shenar for communicating his value of $v_esini$ for BAT99-49 before publication.We also thank the referee, Ian Howarth, for extremely useful correspondence and suggestions. The generosity of the late Paul Newman and the Newman Foundation has made AMNH's participation in SALT possible; MMS gratefully acknowledges that support. S.M.C. acknowledges the South African Astronomical Observatory and the National Research Foundation of South Africa for support during this project. This research made use of Astropy, a community-developed core Python package for Astronomy (Astropy Collaboration, 2013). AFJM is grateful for financial aid from NSRRC (Canada) and FQRNT (Quebec).


\newpage

\label{lastpage}

\end{document}